\documentclass[11pt,a4paper]{article}

\usepackage{amsmath}
\usepackage{amssymb}
\usepackage{amsthm}
\usepackage{a4wide}
\usepackage{longtable}
\usepackage{multicol}
\usepackage{lineno}
\usepackage{relsize}

\everymath{\displaystyle}

\def\it{\textit} 
\def\bf{\textbf} 
 
\def\ul{\underline}

\def\Neg{\mathbf{\neg}}

\def\mc{\mathcal}
\def\mb{\mathbf}

\def\ts{\textstyle}
\def\<_m{<_{\mathrm{m}}}

\newcommand{\be}{\begin{equation}}
\newcommand{\ee}{\end{equation}}
\newcommand{\benum}{\begin{enumerate}}
\newcommand{\eenum}{\end{enumerate}}
\newcommand{\bit}{\begin{itemize}}
\newcommand{\eit}{\end{itemize}}

\newtheorem{thom}{Theorem}
\newtheorem{lemma}[thom]{Lemma}
\newtheorem{rem}[thom]{Remark}

\newtheorem{prob}[thom]{Problem}

\title{Satisfiability formula in CNF for the 
decision version of maximum matching}

\author{Prabhu Manyem\\
University of Newcastle\\
Callaghan NSW Australia}

\begin{document}


\maketitle{\thispagestyle{empty}}

\begin{abstract}
We provide a formula for the lower bound
in the form of $|F| \ge K$, in such a way that the decision version of
unweighted non-bipartite matching can be solved in polynomial time.
~The parameter $K$ can vary from instance to
instance.  We assume that 
the domains, the set of vertices and the set of edges, are ordered.  
To our knowledge, no polynomially solvable satisfiability expression has
been developed for this problem so far, or for that matter, for any
decision problem derived from optimization.  Hence for such problems,
this opens up a new approach to solving them.
\end{abstract}

\bf{Keywords}.
Computational complexity, Satisfiability, Constraint satisfaction.

\bf{AMS classification}.
90C99, 68Q19, 68Q15, 68Q17, 03C13.

\section{Introduction}

We represent decision versions of optimization problems as a conjunction
of a single objective function constraint (OFC) and a set of 
basic feasibility constraints (BFC).
~In other words, such a decision problem $\mc{P}$ can be expressed as

$\mc{P} \equiv$ BFC $\land$ OFC.

The OFC comprises the single constraint 
$|F| \ge K$ for decision problems based on maximization, and
$|F| \le K$ for those based on minimization.
Here $|F|$ is a measure of the size of the objective function.
The OFC 
expresses an upper or lower bound on the size of the set $F$.

The arity of the predicate\footnote{
We may say that a \bf{unary} predicate $F$ is a \bf{subset} of a
\bf{universe} $V$ = \{1, 2, 3, $\cdots$, $n$\}.  Let us illustrate the
terminology with an example.  Suppose 
$V$ = \{1, 2, 3, 4, 5\}, and suppose an algorithm decides that $F(1)$,
$F(3)$ and $F(5)$ are true.  Then we conclude that $F$ = \{1, 3, 5\}, and 
$\Neg F$ (the complement of $F$), is equal to \{2, 4\}.
If the algorithm determines that $F(2)$ and $F(5)$ are true, then we say
that $F$ and $\Neg F$ are equal to \{2, 5\} and \{1, 3, 4\}
respectively.  Thus $F \subseteq V$. ~Similarly if $F$ is binary, then 
$F \subseteq V \times V$.}
 $F$ can vary.  For instance, for a graph
$G = (V,E)$, if $F \subseteq V$, $F$ will be unary (of arity one).
However, if $F \subseteq E$, $F$ will be binary (of arity two).
~In this paper, we will express the lower bound, $|F| \ge K$, 
on a subset $F$ of a given edge set $E$.
The parameter $K$ can vary from instance to instance.

\it{Example}:  Suppose $E$ consists of the edges $i$, $1 \le i \le 7$.
Considering predicate $F$, assume that $F(1)$, $F(3)$, $F(6)$ and $F(7)$
are true, and let the rest, $F(2)$, $F(4)$ and $F(5)$ be false.  In other
words, $F$ = \{1, 3, 6, 7\}. ~If
$K = 3$, then the OFC $|F| \ge K$ is satisfied.  On the
other hand, if $K=5$, the constraint has been violated.

In the case of Matching (Problem \ref{def:matching}), we need to
determine an $F$ ($\subseteq E$) which also satisfies the BFC in Sec.
\ref{sec:matchingBFC}.
For $x \in E$, we say that $F(x)$ is true if and only if $x$ is a
\it{matched} edge.  The decision version of {unweighted non-bipartite
matching} can be formally defined as follows.

\begin{prob}
\bf{Unweighted non-bipartite matching}.
Given an undirected graph $G = (V, E)$ and a non-negative integer $K$, is
there a matching $F$ in $G$, with $F \subseteq E$, such that
\newline
(i) {\em (The BFC)} If two edges $x$ and $y$ meet at a vertex, then at
least one of them should \bf{not} be matched; and
\newline
(ii) {\em (The OFC)} The number $|F|$ of matched edges is at least $K$?
\newline
Assume that (i) $G$ has no self-loops; 
(ii) $G$ is connected;
(iii) $G$ has at most one edge between any pair of vertices;
and (iv) $K \ge 2$.
$\hfill \Box$
\label{def:matching}
\end{prob}
We let $n = |V|$ and $m = |E|$.

\subsection{Significance: Decision problems derived from optimization}\label{sec:significance}

To our knowledge, no polynomially solvable satisfiability expression has
been developed for the decision version of {unweighted non-bipartite
matching} so far, or for that matter, for any decision problem
derived from optimization.

Hence for such problems, this opens up a new approach to solving them.

\subsection{Notation and List of Predicates Used}\label{sec:notation}

Given two integers $a$ and $b$ where $a < b$, 
the set $\big[a, \, b \big]$ within square brackets 
represents the set of integers
\{$a$, $a+1$, $\cdots$, $b-1$, $b$\}.

We write ``$\forall xyz"$ as shorthand for 
"$\forall x \forall y \forall z"$.

For ease of readability, we sometimes write ``edge $x$" in place of
``edge $e_x$" (subscripts and superscripts are harder to read).
We will use $F(x)$ as shorthand for $F(e_x)$ and 
use $L(i, \, x)$ as shorthand for $L(i, \, e_x)$.

There are two types of predicates, the \it{knowns} and the \it{unknowns}.

The \it{knowns} can be determined from the graph input $G = (V,E)$,
or
from the particular value that a universal quantifier can assume in a
certain clause (we use $s$, $t$, $x$ and $y$ for universal quantifiers).

The \it{unknowns}, called ``decision variables" in optimization
terminology, are the ones to be determined by some SAT (satisfiability)
solver.

\begin{rem}
Bounds, such as the upper/lower bounds on the number of positive/negative
literals in a clause, apply only to the \it{unknown} predicates, not the 
\it{known} ones.
$\hfill \Box$
\label{rem:boundsCondition}
\end{rem}
For example, the Horn condition (that the number of positive literals in
a clause must be at most one) applies only to the unknown predicates.

\subsubsection{The knowns}

(a) 
Let's assume that the edges in $E$ are ordered in a certain sequence.
Since $|E| = m$, we can write this sequence as the edges 
$e_1$, $e_2$, $\cdots$, $e_m$, in that order.

Given the ordered sequence $E$, the first and last edges in $E$ are $e_1$
and $e_m$ respectively.

%
 
(b) Define another set $N$, also based on $E$, as follows:

$N$ = \{0, 1, 2, $\cdots$, $m$\} = $\big[0, \, m \big]$.

The difference is that $E$ is a set of edges, whereas $N$ is a set of
non-negative integers.  $N$ is an index set (a set of indices).

(c) The truth values of relations such as $(x < y)$ and $(s > t)$
can be determined from the particular value(s) assumed by $x$, $y$, $s$
and $t$ in a certain clause.

\subsubsection{The unknowns}

(e) Unary $F$, used in Def.  \ref{def:matching}.

(f) Binary $L$, described below.

Define a binary relation $L(i, \, x)$ such that $i \in N$ and $x \in E$.
Examples will be provided in Sec. \ref{sec:combinedSystem}.

We do not need $L$ to be a function; for any $i \in N$, it is possible
that both $L(i, \, x)$ and $L(i, \, y)$ are true when $x \not= y$.
However, we require the inverse of $L$ to be a function; which means that
in $L(i, \, x)$, given $x \in E$, it reverse maps to a unique $i \in N$.

$L$ scans the entire domain $E$ from edge $1$ to edge $m$. As it does so,
it keeps a running count of the number of elements in $E$ that are also
members of $F$ (that is, it keeps a running count of the number of
matched edges).

If edge 1 is matched, then it is assigned index one; otherwise it is
assigned index zero.  From this point onwards the recursion formulas
take over.

Here is the basic idea behind our recursion formulas.  Assume that an edge
$x \in E$ is assigned an index $i \in N$, that is, let $L(i, \, x)$ be true.
~Then, for the next edge $(x+1)$, if it is a matched edge, we increment
the index to $(i+1)$, that is, set $L(i+1, \,  x+1)$ to true.
On the other hand, if $(x+1)$ is unmatched, we leave the index unchanged;
that is, set the index for edge $(x+1)$ to $i$ (which means setting $L(i,
\, x+1)$ to true).

If edge $x \in E$ is \it{mapped} to an index $i \in N$, we say that:
\newline
(i) Edge $x$ has been \it{assigned} index $i$; 
\newline
(ii) $L(i, \, x)$ is true; and
\newline
(iii) The number of matched edges in the set 
\{$e_j ~|~ j \in [1, \, x]$\} is equal to $i$.

\subsection{Directed bipartite graph}\label{sec:bipartite}

It will be helpful to look at the relation $L$ as a directed bipartite graph 
$G_b = (N, E, L)$.  The
elements of $N$ are placed on the left in a vertical column, with
vertex ``0" at the bottom and vertex ``$m$" at the top.  Similarly on the
right side, we have vertices from $E$ in a column with vertex ``$e_1$" at
the bottom and ``$e_m$" at the top.

This is an \it{output} graph, that is, it represents a \it{solution}
obtained after we run the algorithm.

Note that $N$ has one more member than $E$.
~For $i \in [1, \, m]$, place vertices $i \in N$ and $e_i \in E$ at the same
horizontal level. Then vertex $0 \in N$ is one level below the vertices
$1 \in N$ and $e_1 \in E$.


The directed edges in $G_b$ go from left to right (from $N$ to $E$). ~For
a vertex $i \in N$, its out-degree is given by out-degree($i$) $\ge 0$.
However, for a vertex $e_s \in E$, we will require its in-degree to be
one (exactly one arc arrives at $e_s$).

Let us say that an index $i \in N$ is \it{active} if $i$ it is assigned
at least one edge in $E$ and \it{inactive} otherwise.  Recall that the
purpose of the indices is to maintain a running count of the number of
edges matched.  Hence:
\begin{rem}
If index $i$ is active, then all indices less than
$i$ (except perhaps zero) should be active.  As for index zero, it will
be active if and only if the first edge ($e_1$) is unmatched.
$\hfill \Box$
\label{rem:activeIndex}
\end{rem}

\section{BFC for Matching}\label{sec:matchingBFC}

~The BFC for Problem \ref{def:matching} can be written as a Horn formula:
\begin{equation}
\phi_{\textsc{M}}  \equiv
\forall x y ~ [E(x) \land E(y) \land W(x, \, y)] \to [\Neg F(x) \lor
\Neg F(y)].
\label{eq:matchingBFC0}
\end{equation}

$E(x)$ is true if and only if $x$ is an edge in $E$.
~$W(x, \, y)$ is true if and only if edges $x$ and $y$ share a common vertex
in $G$.
~$F(x)$ is true if and only if the edge $x$ is matched.
The \it{known} predicates are $E$ and $W$, whereas $F$ is \it{unknown}.

We can rewrite (\ref{eq:matchingBFC0}) as follows:
\begin{equation}
\phi_{\textsc{M}} \equiv \bigwedge_{(x, \, y) \; \in \; \mc{E}}
[\Neg F(x) \lor \Neg F(y)],
\label{eq:matchingBFC1}
\end{equation}
where $\mc{E}$ is the set of all $(x, \, y)$ pairs such that $x$ and
$y$ are edges in $E$, and they share a common vertex.

\section{A combined system of constraints for the OFC and the
BFC}\label{sec:combinedSystem}

We should express the OFC that $|F| \ge K$.

Carsten Sinz \cite{sinz2005} provided a Horn formula, but his formula is
for the upper bound $|F| \le K$.
~Though his formula could be adapted for lower bounds (as he has himself
suggested in his paper), we develop our own formula in this
paper.

An expression for the lower bound appears in \cite{jabbour2013pigeon}.

\subsection{Modifications to L}\label{sec:modifyL}

\begin{rem}
(a) We define that edge $x$ is assigned an index $i$ if and
only if $L(i, \; x)$ is true.

(b) For edge $m$, we will never use variables $L(i, \; m)$
where $i \in [0, \, K-1]$ in any clause.

(c) We will never use variables $L(i, \, j)$ where $i > j$~ in any of our
clauses.
$\hfill \Box$
\label{def:flip}
\end{rem}

\it{Example 1}:
\newline
Let $E = \{e_i ~|~ i \in [1, \;  10]\}$.  Let the set $F$ of \it{matched}
edges be $F$ = \{1, 3, 7, 9\}.  Then the table of mappings would be as
below.  The matched edges are in bold.

\begin{tabular}{|l|c|c|c|c|c|c|c|c|c|c|}
\hline
Index $i$ &  1 &  1 &  2 &  2 &  2 &  2 &  3 &  3 &  4 &  4 \\
\hline
Edge $x$ &  \bf{1} &  2 &  \bf{3} &  4 &  5 &  6 &  \bf{7} &  8 &  \bf{9} &  10 \\
\hline
\end{tabular}

Edge 5 is assigned index 2 because up to that point, two edges (1 and 3)
have been matched.
$L(2, \;  5)$ is true. 
$L(i, \;  5)$ is false for every $i \neq 5$.

Edge 9 is assigned index 4 because up to that point, four edges (1, 3, 7
and 9) have been matched.
$L(4, \;  9)$ is true.  $L(i, \;  9)$ is false for every
$i \neq 4$.

$L(i, \;  10)$ is false for every $i \neq 4$.
As for index zero, $L(0, \;  e)$ is set to false for every edge $e$, 
since no edge is assigned this index.

The question as to whether $L(0, \; 10)$ is true doesn't arise, since this
variable will never be used in our constraints (part (b) of 
Remark \ref{def:flip}).

\it{Example 2}:
\newline
$E = \{e_i ~|~ i \in [1, 10]\}$ and 
$F$ = \{3, 7, 9, 10\} would result in the following mapping table:

\begin{tabular}{|l|c|c|c|c|c|c|c|c|c|c|}
\hline
Index $i$ &  0 &  0 &  1 &  1 &  1 &  1 &  2 &  2 &  3 &  4 \\
\hline
Edge $x$ & 1 & 2 & \bf{3} & 4 & 5 & 6 & \bf{7} & 8 & \bf{9} & \bf{10} \\
\hline
\end{tabular}

Edges 1 and 2 are assigned index zero because up to that point, no edge
has been matched.  We see that 
$L(0, \;  1)$ and $L(0, \; 2)$ are set to true.
$L(0, \;  e)$ will be set to false for every edge $e$ other than 1 and 2.

Again, the issue of whether $L(0, \; 10)$ is true is none of our
concern.

\subsection{Properties of lines L in the output graph $\mb{G_b}$}

Any reference to ``edge $x$" in this section (Sec.
\ref{sec:combinedSystem}) will refer to an edge in the given (input)
graph $G = (V,E)$, \it{not} in the (output) bipartite graph $G_b$.

This subsection is to help the reader understand $L$.
~There are no definitions, constraints or proofs.

\subsubsection{No downward slopes}\label{sec:downSlope}

Recall that $N$, the set of indices, is defined as \{$i~|~0 \le i \le
m$\}, where $m$ is the number of the edges in the given graph $G =
(V,E)$.

In all our constraints, only variables $L(i, \; j)$ where $i \le j$ will be used.
Moving from the left set $N$ to the right set $E$ in $G_b$, the
slopes of the lines in $L$ will always be non-negative.  Thus we will
never use variables of the form $L(i, \; j)$ where $i > j$.

The slope of any line $L(i, \; j)$ is either flat or upward.

If the slope is flat at Level $j$ (that is, if edge $j$ is assigned
index $j$), this means that for every level below $j$, that is, for every
$p \in [1,~j-1]$, 
\begin{enumerate}
\item
Edge $p$ is matched (that is, $F_p$ is true); and
\item
At Level $p$, the slope is flat ($L(p, \; p)$ is true).
\end{enumerate}

A flat slope at Level $j \in [1, ~m]$ would mean that $L(j, \; j)$
is true.
As we move up from any edge $j$ to edge $(j+1)$ where $j \in 
E - \{m\}$, the slope will either increase or remain the same (never
decrease).

\subsubsection{Parallel lines (with equal slopes)}

Let the slopes of two lines in $G_b$ be parallel.
In other words, assume that edge $x$ is assigned
index $i$, edge $y$ is assigned $j$, and $(x-i) = (j-y)$.
This implies that every edge in the interval between them is matched,
including $y$ (but we are unable to say whether $x$ is matched).

That is, if $L(i, \; x)$ and $L(j, \; y)$ are true for $1 \le i < j \le m$,
$1 \le x < y \le m$ and $(x-i) = (j-y)$, then every edge in the
interval $[(x+1),~ y]$ is matched.

The parallel lines system is applied in Constraint 
(\ref{eq:findMatchDiffIndex}).

We can now begin writing down the constraints one by one.

\subsection{The cardinality constraint}\label{sec:cardinality}

The cardinality condition that the ``index of edge $m$ should be at
least $K$" could be written as a conjunction of $K$ unit clauses:
\begin{equation}
\eta(1) \equiv 
\bigwedge_{\ts i \in [0,\; K-1]}  \Neg L(i, \, m).
\label{eq:indexMcanAssume}
\end{equation}
(A unit clause is a clause with only one literal.)

However, we can do something simpler.  We will never use these variables 
$L(i, \, m)$, where $i \in [0,\; K-1]$, in any of our clauses.  Hence
constraint (\ref{eq:indexMcanAssume}) is unnecessary and it can be
ignored.

Later in Sec. \ref{sec:atLeastOneEdge}, we will ensure that every edge is
assigned at least one index.  This and the uniqueness condition below
will ensure that edge $m$ is
assigned exactly one index in the range $[K, \, m]$.

\subsection{Uniqueness of mapping for edges}\label{sec:uniqueness}

We will ensure that every edge $x \in E$ is assigned a unique index 
$i \in N$. 

Considering $L(i, \; x)$ and $L(j, \; x)$, assume w.l.o.g. that
$i < j$.
For edge $x$ where $x \in [1, ~m]$:
\begin{equation}
\eta(2) ~\equiv~ \Neg L(i, \;  x) \lor \Neg L(j, \;  x).
\label{eq:uniqueness1}
\end{equation}
Recall that we neither need nor use variables $L(i, \; x)$ where $i > x$.

As a special case, for the first edge $e_1$,
the only possibilities for $i$ and $j$  are $i = 0$ and $j = 1$.
If edge $e_1$ is matched, then we need $L(1, \; 1)$ to be true,
otherwise we need $L(0, \; 1)$ to be true.
We state that the two variables cannot be simultaneously true:
\begin{equation}
\eta(3) ~\equiv ~\equiv~ \Neg L(1, \;  1) \lor \Neg L(0, \;  1).
\label{eq:uniqueEdge1}
\end{equation}

\subsection{Main mapping L: forward recursion}\label{sec:mainMap}

This is forward recursion; recursion from the ground up.

\subsubsection{The different cases for different edges}\label{sec:mainMapBasic}

(a) If edge $e_1$ is matched, then $L(1, \; 1)$ should be set as true, otherwise
$L(0, \; 1)$ should be set as true:
\begin{equation}
\eta(5) \equiv L(1, \;  1) \lor L(0, \;  1).
\label{eq:mapEdge1}
\end{equation}
Variable $L(i, \; 1)$ is undefined for $i \ge 2$.  Thus the choice of indices
in $N$ for edge $e_1$ is restricted to the set \{0, 1\}.

(c) For $x \in [1, ~m-1]$ and $i \in [0,\, x]$,
if edge $x$ is assigned a non-zero index $i$, then
the next edge $(x+1)$ is assigned either $i$ or $(i+1)$:
\begin{equation}
\begin{aligned}
\eta(8) ~\equiv~ & L(i, \; x) \to \big[ L(i, \;  x+1) \lor L(i+1, \;  x+1) \big] \\
        ~\equiv~ & \Neg L(i, \; x) \lor L(i, \;  x+1) \lor L(i+1, \;  x+1).
\end{aligned}
\label{eq:mapEdgeXnot0}
\end{equation}
But let us tighten this.
If $x$ is assigned index $i$, then let us ensure that edge $(x+1)$ will
be assigned \it{only} indices $i$ or $(i+1)$:
\begin{equation}
\begin{aligned}
\eta(9) ~\equiv~ & L(i, \; x) \to \left[ 
\bigwedge_{\ts j \in [0, \; x+1],\; j \notin \{i, \, i+1\}} 
               \Neg L(j, \; x+1) \right] \\ 
  & \\
~\equiv~ & 
\bigwedge_{\ts j \in [0, \;  x+1], \; j \notin \{i, \, i+1\}} 
\big[ \Neg L(i, \; x) \lor \Neg L(j, \;  x+1) \big].
\end{aligned}
\label{eq:tightenEdgeXnot0}
\end{equation}

When $i = K-1$, since $L(K-1, \, m)$ is undefined, Equation
(\ref{eq:mapEdgeXnot0})
becomes as follows, forcing edge $m$ to be matched (and hence assigned
index $K$):
\begin{equation}
\eta(10b)  ~\equiv~ L(K-1, \,m-1) \to L(K, \, m)
      ~\equiv~  \Neg L(K-1, \, m-1) \lor L(K, \, m).
\label{eq:mapEdgeM2}
\end{equation}

Recall that variables $L(j, \; m)$ are undefined for $j \in [0, \, K-1]$
(part (b) of Remark \ref{def:flip}).

Thus for every edge in $E$, we have restricted its choice of indices in
$N$ to at most two.  The uniqueness conditions in Sec. \ref{sec:uniqueness}
and the matching definitions in Sec. \ref{sec:defineMatch}
will enforce that exactly one of these two indices will be chosen for
every edge.

\subsubsection{At least one index for every
edge}\label{sec:atLeastOneEdge}

It is easy to show that every edge in $E$ is assigned an index $i$ in
$N$.

The first edge $e_1$ is assigned an index (zero or one), as per
(\ref{eq:mapEdge1}).  From then on, the recursion equations 
(\ref{eq:mapEdgeXnot0}) take over and assign an index to
every edge.


\begin{lemma}
The recursion defined by (\ref{eq:mapEdge1}) and (\ref{eq:mapEdgeXnot0})
ensure that every edge in $E$ is assigned at least one index in $N$.
$\hfill \Box$
\label{lemma:atLeastOneIndex}
\end{lemma}

\subsection{Tightly coupled systems}

The three events $L(i, \, x)$, $F(x+1)$ and
$L(i+1, \, x+1)$ form a tightly coupled system for $i \in [1, \, m-1]$
and $x \in [1, \, m-1]$.  If two of them are true, then so is the third.

Similarly, the three events $L(i, \, x)$, $\Neg F(x+1)$ and $L(i, \, x+1)$
form a tightly coupled system for the same values of $i$ and $x$ as above.

The three sections, \ref{sec:defineMatch}, \ref{sec:forwardAgain} and
\ref{sec:backwardRec}, each define the truth of the third event in terms
of the other two.

(a) As we go up from edge $x$ to $(x+1)$, the index is incremented if and
only if the succeeding edge $(x+1)$ is matched.
(Sec. \ref{sec:forwardAgain})

(b) Similarly, as we move downwards from edge $(x+1)$ to $x$, the index
to which $x$ gets assigned depends on whether $(x+1)$ is matched.
(Sec. \ref{sec:backwardRec})

(c) Is edge $x$ matched?  This depends on whether $(x-1)$ is assigned
the same index as $x$, or the one below. (Sec. \ref{sec:defineMatch})

The constraints in Sections \ref{sec:defineMatch} and 
\ref{sec:forwardAgain} together enforce the following condition: 
If edge $x$ is assigned index $i$, 
 then exactly one of the following is true about edge $(x+1)$:

Either
(i) it is assigned index $(i+1)$ and it is matched,
\newline
or
(ii) it is assigned index $i$ and it is unmatched.

\subsection{Defining whether an edge is matched}\label{sec:defineMatch}

In Sec. \ref{sec:mainMap}, we were able to
restrict the choice of indices for every edge to at most two.  In this
subsection, by defining a matching $F(x)$ for every edge $x$, we can
restrict this choice to one (since $F(x)$ can be true or false, not
both).

We now define which edges are matched and which ones are not.

We define an edge $x$ to be \it{matched} if $F(x)$ is true, and
\it{unmatched} (or \it{not matched}) if $F(x)$ is false.

(a)
From (\ref{eq:uniqueEdge1}) and (\ref{eq:mapEdge1}), we can state the
conditions for the first edge, edge $e_1$:
\begin{equation}
\begin{aligned}
\eta(12) ~\equiv~ & L(1, \; 1) \to F(1) & \equiv~ \Neg L(1, \; 1) \lor F(1) \\ 
\eta(13) ~\equiv~ & L(0, \; 1) \to \Neg F(1) & ~\equiv~ \Neg L(0, \; 1) \lor \Neg F(1).
\end{aligned}
\label{eq:findMatch1}
\end{equation}

(b) If an edge $x$ is assigned index zero, clearly it should not be matched.
For edge $x$, where $x \in [2, ~m-1]$, we define:
\begin{equation}
\eta(14) ~\equiv~ L(0, \; x) \to \Neg F(x) ~\equiv~ \Neg L(0, \; x) \lor \Neg F(x).
\label{eq:MatchIndex0}
\end{equation}

(c) For $x \ge 2$, if the index of edge $x$ is zero, then every edge $p$
in the set $[1, ~x-1]$ will also have an index of zero (and hence be
unmatched):
\begin{equation}
\begin{aligned}
\eta(15) ~\equiv~ &  L(0, \; x) \to  L(0, \, x-1) \\ 
~\equiv~ & \Neg L(0, \; x) \lor L(0, \, x-1).
\end{aligned}
\label{eq:findMatchIndex0}
\end{equation}

(d) Edge $x$, where $x \in [2, ~m-1]$, is matched if edges $x$ and
$(x-1)$ are assigned different indices in $N$:
\begin{equation}
\begin{aligned}
(\text{For}~ i \ge 1)~~
\eta(16) ~\equiv~ & \big[ L(i, \; x) \land L(i-1, \,  x-1) \big] \to F(x) \\ 
         ~\equiv~ & \Neg L(i, \; x) \lor \Neg L(i-1, \,  x-1) \lor F(x).
\end{aligned}
\label{eq:findMatchDiffIndex}
\end{equation}

(e) Edge $x$, where $x \in [2, ~m-1]$, is \it{not} matched if edges $x$
and $(x-1)$ are assigned the same index in $N$:
\begin{equation}
\begin{aligned}
\eta(18) ~\equiv~ & [L(i, \; x) \land L(i, \,  x-1)] \to \Neg F(x) \\
       ~\equiv~ & \Neg L(i, \; x) \lor \Neg L(i, \,  x-1) \lor \Neg F(x).
\end{aligned}
\label{eq:findMatchSameIndex}
\end{equation}

Obviously an algorithm will set $F(m)$ to be either true or false, not
both.  If it is chosen to be true, then the unique index for edge $m$
will be $(i+1)$, otherwise it is $i$.

\subsection{Forward recursion revisited}\label{sec:forwardAgain}

Having defined matching $F(x)$ for every edge $x$, we can rewrite forward
recursion in terms of matching.
As we go up from edge $x$ to $(x+1)$, the index $i$ is incremented if and
only if the succeeding edge $(x+1)$ is matched.

(a) For $x \in [1, ~m-1]$ and $i \in [1,x]$,
if edge $x$ is assigned a non-zero index $i$, then
the next edge $(x+1)$ is assigned either $i$ or $(i+1)$, depending on
the value of $F(x+1)$:
\begin{equation}
\begin{aligned}
\eta(8b) ~\equiv~ & \big[ L(i, \, x) \land F(x+1) \big] \to L(i+1, \,  x+1) \\
        ~\equiv~ & \Neg L(i, \, x) \lor \Neg F(x+1) \lor L(i+1, \,  x+1). \\
\eta(8c) ~\equiv~ & \big[ L(i, \, x) \land \Neg F(x+1) \big] \to L(i, \,  x+1) \\
        ~\equiv~ & \Neg L(i, \, x) \lor F(x+1) \lor L(i, \,  x+1).
\end{aligned}
\label{eq:forwardAgain}
\end{equation}

If $L(K-1, \, m-1)$ is true, then edge $m$ must be matched:
\begin{equation}
\begin{aligned}
\eta(10d) ~\equiv~ & L(K-1, \, m-1) \to \big[F(m) \land L(K, \,  m) \big] \\
        ~\equiv~ & \big[ \Neg L(K-1, \, m-1) \lor F(m) \big] \land
           \big[ \Neg L(K-1, \, m-1) \lor L(K, \, m) \big].
\end{aligned}
\label{eq:forwardAgain4}
\end{equation}


The result in Lemma \ref{lemma:atLeastOneIndex} could also have been
derived from the three constraints in this subsection alone.

\subsection{Backward recursion}\label{sec:backwardRec}

If edge $x$ is assigned index $i$, how can we be sure that the previous
edge $(x-1)$ is assigned either $i$ or $(i-1)$?  We ensure this property
in this subsection.

This is recursion from top to bottom (edge $m$ to edge 1).

If edge $x$ is assigned index $i$, then the previous edge $(x-1)$ should
be assigned either $i$ or $(i-1)$, depending on the status of edge $x$
(whether it is matched).

The case when $i = 0$ was treated in (\ref{eq:MatchIndex0}) and 
(\ref{eq:findMatchIndex0}).

(a) For edge $x$ where $x \in [2, ~m]$ and index $i \in [1, \,  m-1]$:
\begin{equation}
\begin{aligned}
\eta(22) ~\equiv~ & \big[ L(i, \; x) \land F(x) \big] \to L(i-1, \,  x-1) 
         ~\equiv~   \Neg L(i, \; x) \lor \Neg F(x) \lor L(i-1, \,  x-1). \\ 
\eta(23) ~\equiv~ & \big[ L(i, \; x) \land \Neg F(x) \big] \to L(i, \, x-1) 
         ~\equiv~   \Neg L(i, \; x) \lor F(x) \lor L(i, \, x-1).
\end{aligned}
\label{eq:backwardEasyCase}
\end{equation}

(b) If $L$ is flat at Level $p$, then $L$ is flat at every level below
$p$:
\begin{equation}
\eta(24) ~\equiv~  L(p, \, p) \to L(p-1, \, p-1)
 ~\equiv~  \Neg L(p, \, p) \lor L(p-1, \, p-1), ~ p \in [2, \, m].
\label{eq:flatAtEveryLevel}
\end{equation}
Combining the above with (\ref{eq:findMatchDiffIndex}) in Sec.
\ref{sec:defineMatch}, this means that all edges in the range $[1, \, p]$
are matched.

Next, we fit the BFC in Sec. \ref{sec:matchingBFC} into mapping $L$.

\subsection{Embedding the BFC into the Main mapping}\label{sec:bfcFit}

(The ideas in this section are due to L. Brankovic.  I thank her for
this.)

Consider the case when an edge $x$ is assigned an index $i$.   From Sec.
\ref{sec:mainMap}, we know that if the
next edge $(x+1)$ is matched, its index is set to $(i+1)$; otherwise
its index remains the same as that of edge $x$, which is $i$.

Now consider two different edges $x$ and $y$.  Assume w.l.o.g. that 
$x < y$.  Assume that $W(x, \; y)$ is true, that is, edges $x$ and $y$ meet
at a common vertex in $V$.
~Then at least one of these two edges should be unmatched.  In
other words, either
\newline
(i) Edges $x$ and $(x-1)$ should share the same index; or
\newline
(ii) Edges $y$ and $(y-1)$ should share the same index; or
\newline
(iii) Both (i) and (ii).

This is expressed as below.
Assume that edges $x$ and $y$ are assigned indices $i$ and $j$
respectively.
In Cases (a) and (b), the satisfaction of the BFC is immediately clear.
Only for Cases (c) and (d), we are required to write the conditions
explicitly.

(a) If $i = j$:

This means that all edges in the set \{$x+1$, $\cdots$, $y$\} are
assigned the same index.
Constraint (\ref{eq:findMatchSameIndex}) implies that all these edges,
including $y$, are unmatched.  Hence the BFC is satisfied.

(b) If $i = 0$ and $i < j$:

Constraint (\ref{eq:findMatchIndex0}) implies that edge $x$ is unmatched,
hence satisfying the BFC regardless of whether $y$ is matched.
~(Also, for $p \le x$, every edge $p$ is unmatched.)

(c) If $0 < i < j$ and $y \le m$:
\begin{equation}
\begin{aligned}
\eta(25) ~\equiv~ & \big[ L(i, \; x) \land L(j, \; y) \big] \to [L(i, \, x-1)
                     \lor L(j, \, y-1)] \\ 
         ~\equiv~ & \Neg L(i, \; x) \lor \Neg L(j,y) \lor
                  L(i, \, x-1) \lor L(j, \, y-1).
\end{aligned}
\label{eq:fitBFC1}
\end{equation}
Applying (\ref{eq:findMatchSameIndex}), we can also express this as:
\begin{equation}
\begin{aligned}
\eta(26) ~\equiv~ & \big[ L(i, \; x) \land L(j, \; y) \big] \to 
                    \big[ \Neg F(x) \lor \Neg F(y) \big] 
\\ 
         ~\equiv~ & \Neg L(i, \; x) \lor \Neg L(j, \; y) \lor \Neg F(x) \lor \Neg F(y).
\end{aligned}
\end{equation}
A special case arises when $i = x$:

If $i = x$, that is, if $L$ is flat at Level $x$, then we know from 
(\ref{eq:findMatchDiffIndex}) and (\ref{eq:flatAtEveryLevel})
that edge $x$ is a matched edge, and that $L(x, \; x)$ and $F(x)$ are true.
Then $\Neg F(x)$ is false and hence can be removed from the equation.
Also, there is no such variable $L(x,\; x-1)$ (from the last part of Def.
\ref{def:flip}).

Since $x$ is matched, edge $y$ should be \it{unmatched}.
The two constraints above simplify to 

$\Neg L(x, \; x) \lor \Neg L(j, \; y) \lor L(j,\, y-1)$ ~and~ 
$\Neg L(x, \; x) \lor \Neg L(j, \; y) \lor \Neg F(y)$

respectively.

But what if $j = y$?  Then the slope is flat at Level $y$ and hence edge
$y$ is matched.  We should apply Eq. (\ref{eq:flatAtEveryLevel}).  This
would mean that the slope is flat at every level below $y$, including
level $x$, hence edge $x$ is also matched.  Thus both edges $x$ and $y$
will have to be matched.  Hence for such $(x, \; y)$ pairs of edges, we
enforce that
\begin{equation}
\eta(27) ~\equiv~ \Neg L(y, \, y).
\end{equation}

(\it{This subsection is probably unnecessary, since the BFC in Sec.
\ref{sec:matchingBFC} has negative literals in every clause}.)

\section{Conclusion}

Every clause in the system of constraints in Sec.
\ref{sec:combinedSystem} has a polynomial number of literals.  
In fact, every clause has at most four literals.
Every clause except (\ref{eq:mapEdge1}) has at least one negative
literal.
The number of constraints is also polynomial in $n$.

If (\ref{eq:mapEdge1}) can also be converted to a clause with at least
one negative literal, then from \cite{schaefer}, it follows that:
\begin{thom}
The system of constraints 
in Sec.  \ref{sec:combinedSystem}, from 
(\ref{eq:uniqueness1}) onwards, 
can be solved in time polynomial in $n$.
\end{thom}


\subsection{For future versions of this paper: Possible further
restrictions to L}

We could restrict $L$ further, based on the cardinality constraint.

\ul{Restrictions for $E$:}
\newline
For every edge $x$, we could define a range of indices to which $x$ could
be assigned and leave other indices undefined (for $x$).

To start, for edge $(m-1)$, its defined range of indices will be
$[K-1, \, m-1]$.  Hence $L(i, \, x)$ where $i \notin [K-1, \, m-1]$ can be
left undefined (and thus not used in any clause).

In general, the defined range of indices for edge $(m-i)$, for $i \in
[0, \, K]$ would be $[K-i, \, m-i]$.

\ul{Restrictions for $N$:}
\newline
Since we require that $|F| \ge K$, at most $m-K$ edges can be assigned an
index of zero. Hence this index can be assigned to edges in the range
$[1, \, m-K]$.

In general, index $i$, where $i \in [0, \, K]$, can be restricted to
edges in the range $[i, \, m-K+i]$.

\subsection{Further Notes}

%

(a) We have enforced the uniqueness condition for each edge in two ways;
in Sections \ref{sec:uniqueness} and \ref{sec:mainMap}.  I suppose that
one of these could be dropped (perhaps the one in Sec.
\ref{sec:uniqueness}). 

(b) Embedding the BFC into mapping $L$ (Sec. \ref{sec:bfcFit}) is
probably unnecessary, since the BFC in 
Sec. \ref{sec:matchingBFC} has at least one negative literal in every
clause.

(c) I think the 3 sections, BFC (Sec. 2), "Forward recursion revisited"
(Sec. \ref{sec:forwardAgain}) and "Defining Matching" (Sec.
\ref{sec:defineMatch}) are sufficient to define the problem.

\subsection{Acknowledgement}

I am grateful to Ljiljana Brankovic for contributing Subsection
\ref{sec:bfcFit}.


\begin{thebibliography}{1}

\bibitem{jabbour2013pigeon}
Said Jabbour, Lakhdar Sais and Yakoub Salhi.
\newblock A Pigeon-Hole Based Encoding of Cardinality Constraints.
\newblock In {\em International Symposium on Artificial Intelligence and
Mathematics, {ISAIM} 2014}, Fort Lauderdale, FL, USA, January 6-8, 2014.

\bibitem{schaefer}
Thomas~J. Schaefer.
\newblock The complexity of satisfiability problems.
\newblock In {\em Proceedings of the Tenth Annual ACM Symposium on Theory of
  Computing}, STOC '78, pages 216--226, New York, NY, USA, 1978. ACM.

\bibitem{sinz2005}
Carsten Sinz.
\newblock Towards an optimal CNF encoding of boolean cardinality constraints.
\newblock In Peter van Beek, editor, {\em Principles and Practice of Constraint
  Programming - CP 2005}, volume 3709 of {\em Lecture Notes in Computer
  Science}, pages 827--831. Springer Berlin Heidelberg, 2005.

\end{thebibliography}
\end{document}